\documentstyle[prl,aps]{revtex}
\begin{document}

\input{epsf}
\draft

\twocolumn[\hsize\textwidth\columnwidth\hsize\csname @twocolumnfalse\endcsname
  
\title{Real-Space Renormalization Group Study of the
Two-dimensional Blume-Capel Model with a Random Crystal Field\cite{cnpq}}

\author{ \sc { N.\ S.\ Branco }} 
\address{Departamento de F\'{\i}sica - Universidade Federal de Santa
Catarina\\ 
88040-900, Florian\'opolis, SC  - Brazil; e-mail:
nsbranco@fsc.ufsc.br \\}
\author{ \sc { Beatriz M.\ Boechat }}
\address{Departamento de F\'{\i}sica - Universidade Federal Fluminense\\
24210-130, Niter\'oi , RJ  - Brazil;   e-mail: bmbp@if.uff.br \\}

\date{\today}

\maketitle

\begin{abstract}

 	The phase-diagram of the two-dimensional Blume-Capel model 
with a random
crystal field is investigated within the framework of a real-space 
renormalization group approximation. 
Our results suggest that, for any amount of randomness,
the model exhibits a line of
Ising-like continuous transitions, as in the pure model, but
no first-order transition. At zero temperature the transition is
also continuous, but not in the same universality class as the
Ising model. In this limit, the attractor (in the renormalization 
group sense) is the percolation fixed point of the
site diluted spin-1/2 Ising model. The results we found are in qualitative 
agreement with general predictions made by Berker and Hui on the critical 
behaviour of random models.

\end{abstract}

 
\vskip2pc]

\section{Introduction}
                  
     The Blume-Capel (BC) model is a spin-1 Ising model, originally 
proposed to study first-order magnetic phase transitions \cite{Blu}: 
its phase-diagram presents a line of continuous
transitions and a line of first-order transitions, separated by a
tricritical point. The Hamiltonian of the model is given by
\begin{equation}
   {\cal  H}_{BC} = - J \sum_{<i,j>} S_i S_j + \Delta \sum_i  S_i^2, 
   \label{eq:hamil} 
\end{equation}
where the first sum is over all nearest-neighbor 
pairs on a lattice and the last one is over all sites, $J$ is the
exchange constant, $\Delta$ is the crystal field and $S_i=\pm 1,0$.
Later, a generalization of the BC model was introduced, 
the Blume-Emery-Griffiths (BEG) model \cite{BEG}: it has been used to study
a rich variety of physical systems, among them $^3$He-$^4$He mixtures.
Its Hamiltonian reads: 
${\cal H} = {\cal  H}_{BC} - K \sum_{<i,j>} S_i^2 S_j^2$ 
and the parameters $J$, $K$ and $\Delta$ were originally
related to the energy interactions between the constituents of
the system. In $^3$He-$^4$He mixtures,
the state $S=0$ represents a $^3$He atom while $^4$He atoms are
denoted by $S=\pm 1$ states: the superfluid transition 
corresponds to the symetry breaking between the $\pm 1$ states.

	More recently, the critical behavior of $^3$He-$^4$He mixtures
in random media (more precisely, in aerogel) has been modeled by a BC
model with a random crystal field (RFBC).
The presence of the porous media is taken into
account by the introduction of a site-dependent crystal field, which 
follows the probability distribution
\begin{equation}
   {\cal P}(\Delta_i) = p \; \delta(\Delta_i-\Delta_1) +
(1-p) \; \delta(\Delta_i-\Delta_2) , \label{dist}
\end{equation}
where $\Delta_1$ is the field at the pore-grain interface
and $\Delta_2$ is a bulk field which controls the concentration
of $^3$He atoms \cite{mar1,mar2}. More precisely, the {\it BEG model} 
has been used to describe $^3$He-$^4$He mixtures, and the
biquadratic exchange parameter, $K$, is related to 
the interaction energy between $^{\alpha}$He-$^{\beta}$He atoms,
$V_{\alpha\beta}$, through $V_{33} + V_{44} - 2 V_{34}$.
Since $V_{\alpha\beta}$ is nearly independent of $\alpha$ and $\beta$, 
one can assume that $K$ is zero, regaining the Blume-Capel model.

	From the theoretical point of view, the presence of randomness
may affect the critical behavior of systems in a drastic way.
Random bonds \cite{Har} and random fields \cite{Imry} effects
on phase transitions have been studied for a long time. 
Briefly, the effect of random fields on multicritical phase
diagrams is the following: the presence of an infinitesimal ammount
of randomness eliminates non-symmetry-breaking first-order 
transitions and replaces
symmetry-breaking first-order transitions by continous ones
in two dimensions ($d=2$), while for $d>2$ tricritical points and critical 
end points are depressed in temperature and first-order phase
transitions are supressed only at a finite amount of the disorder
\cite{berker}. Whether the
first-order transition in two dimensions is replaced by a continuous
transition in the same universality class as the spin-1/2 Ising model, or
otherwise, is still an open problem \cite{mar2}.

	To the best of our knowledge, theoretical works with
 the RFBC model have used some sort of mean-field-like approximation 
\cite{mar1,mar2,mf1,mf2}. These approximations describe correctly the behavior
of high-dimensional systems and, even in those cases, a meaningful discussion
on universality classes is not possible. The so called effective-field
approximation \cite{mf2}, for instance,
cannot describe first-order phase transitions; on the other hand, standard
mean-field approaches (which assume that each spin interacts with all other
spins in the system) do not lead to a correct discussion of percolation
effects.

	In the present work, we employ a  real-space renormalization group (RSRG)
approximation to discuss the two-dimensional RFBC model. 
Our approximation takes into account spin correlations at all levels, 
and allows for the discussion of first-order transitions \cite{fisher},
universality classes, multicritical points \cite{berker1}, etc.
The crystal field probability distribution we chose is slightly different
from Eq.\ (\ref{dist}), namely
\begin{equation}
   {\cal P}(\Delta_i) = p \; \delta(\Delta_i+\Delta) +
    (1-p) \; \delta(\Delta_i-\Delta) . \label{distri} 
\end{equation}
We believe that the important
physical ingredient lies on the presence of randomness and not on the exact
form of the probability distribution. In fact, we performed a mean-field
calculation using Eq.\ (\ref{distri}) as the probability distribution
and the results are qualitatively equivalent to those obtained using
different distributions and the same mean-field approximation.

	The remainder of this paper is organized as follows. In section
II we outline the formalism and discuss some technical points, in section III
we present the results, and in the last section we summarize our main
conclusions.  

\section{Formalism}

\begin{figure}
\epsfxsize=6.5cm
\begin{center}
\leavevmode
\epsffile{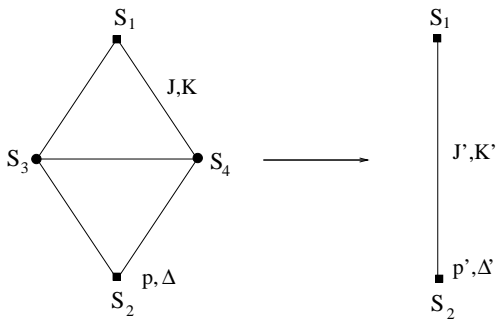}
\caption{Construction of a hierarchical lattice  adequate
to simulate the square lattice. $S_1$ and $S_2$ 
denote terminal spins while $S_3$ and $S_4$ denote internal spins. 
The original lattice is depicted on the left-hand side of the figure, 
with parameters $p, J, K,$ and $\Delta$; after doing a partial trace over 
spins $S_3$ and $S_4$ we are left with the ``new'' lattice (right-hand 
side), with renormalized parameters $p', J', K'$ and $\Delta'$. Note that
the construction of the hierarchical lattice is done by reversing the
length scale transformation.}
\label{cell} 
\end{center}
\end{figure}

 We approximate the Bravais
lattice, which in our case, is the square lattice, by an appropriate
hierarchical lattice. We chose one of the simplest cells, depicted in
Fig.~\ref{cell}; albeit its simplicity, this cell has been used with
success in the study of many ferromagnetic systems. We note that the results
obtained are {\it exact} on the chosen hierarchical lattice but only
approximate on the square lattice. In particular, one does not
expect to obtain results as precise as  those from
Monte Carlo simulations or conformal invariance arguments. 
Nevertheless, universality classes and the order of the transitions are
very well described by RSRG approximations, particularly in two dimensions.

     We then impose that the correlation function between the two terminal 
sites of the original and renormalized graphs are preserved \cite{tsallis}:
\begin{equation}
  \exp (- \beta {\cal H}_{12}) = Tr \; 
\exp (- \beta {\cal H}_{1,2,3,4}), 
\end{equation} 
where $Tr$ means a partial trace over the internal sites of the
cell ($S_3$ and $S_4$ in Fig.~\ref{cell}).
We rewrite the cell Hamiltonians as a sum of ``bond'' terms
(from now on, the factor  $\beta$ will be absorbed into the 
interaction parameters)
\begin{eqnarray}
{\cal H}_{1,2,3,4}&=& - J (S_1 S_3 + S_1 S_4 + S_3 S_4 + S_2 S_3 +
 S_2 S_4 ) \nonumber \\
 & &- K (S_1^2 S_3^2 + S_1^2 S_4^2 + S_3^2 S_4^2 + S_2^2 S_3^2 + S_2^2 S_4^2 ) 
 \nonumber \\& &+ ( 2 \frac{\Delta_1}{4} S_1^2 + 3 \frac{\Delta_3}{4} S_3^2 + 
 3 \frac{\Delta_4}{4} S_4^2 + 2 \frac{\Delta_4}{4} S_2^2 ) , \label{cell1}
\end{eqnarray} 
where the crystal field $\Delta_i$ follows the probability distribution
given by Eq.\ (\ref{distri}), and
\begin{equation}
   {\cal H}_{12} = - J' S_1 S_2 - K' S_1^2 S_2^2 + 
 (\frac{\Delta'_1}{4} S_1^2 + \frac{\Delta'_2}{4} S_2^2) + G' ,
 \label{cell2}
\end{equation}
where primed quantities are renormalized parameters and $G'$ is
a constant, generated by the renormalization procedure.
We comment below on the renormalized probability distributions.

\begin{figure}
\epsfxsize=6.5cm
\begin{center}
\leavevmode
\epsffile{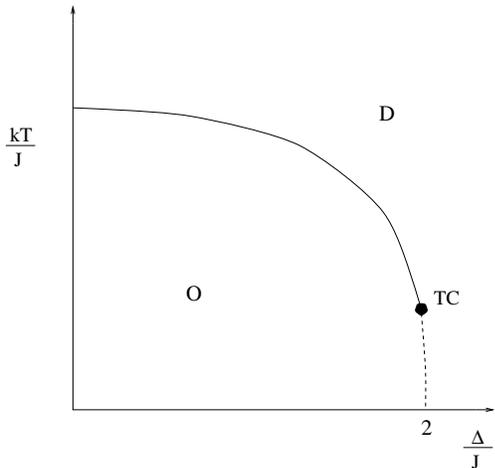}
\caption{Phase diagram of the pure Blume-Capel model, where
$k$ is the Boltzmann constant and $T$ is the temperature. The first order
(dashed) line flows to a zero temperature fixed point, where the
largest eigenvalues for odd and even sectors of the RGT matrix equals
$b^d$,  where 
$b$ is the length-scaling parameter and $d$ is the dimension of the system.
$TC$ is the tricritical point, which flows to 
$J^*=1.51,K^*=0.051$ and $\Delta^*=3.01$ (within the present approximation).
The continuous line, to the left of $TC$, is atracted to the spin-1/2
Ising fixed point. $O$ ($D$) stands for ordered (disordered) phase.}
\label{pure} 
\end{center}
\end{figure}

  Note that this way to write the cell Hamiltonians (Eqs. \ref{cell1}
and \ref{cell2}) is equivalent to  attribute weights to the 
sites in the one-site  (crystal-field
$\Delta$) interaction, according to their coordination number.
This is necessary for finite lattices in order to approximate correctly the
infinite lattice behavior (see, for instance, Ref\cite{dico}). By using
the above procedure we obtain the exact value for the point where both  
ferromagnetic and paramagnetic phases coexist
at zero temperature for the pure ($p=0$) Blume-Capel
model on the square lattice, namely $(\Delta/J)_c=2$ 
(see Fig.~\ref{pure}).

	Some points are worthy stressing at this stage. 
First, we comment on the presence of the biquadratic interaction $K$
in our formalism. Although we are treating
the Blume-Capel model ($K=0$), the parameter $K$ is generated by the 
renormalization procedure and it must be taken into account to follow the
renormalization path. To restrict oneself to a subspace which is not
invariant usually leads to spurious results. Second,
the renormalization procedure will introduce randomness in all
renormalized quantities ($J', K'$ and $\Delta'$). One possible approach
would be to follow the successive renormalized distributions of these
parameters in order to study the phase diagram. We adopted an alternative
way which forces the renormalized distributions to be the same as the
initial ones, but with renormalized parameters, namely,
$ {\cal P}'_{ap}(J) = \delta(J-J')$, ${\cal P}'_{ap}(K) = \delta(K-K')$ and
${\cal P}'_{ap}(\Delta_i) = p' \; \delta(\Delta_i+\Delta') + (1-p') \;
\delta(\Delta_i-\Delta')$. The values of $J'$ and $K'$ are
obtained by imposing that the first moment of the actual distributions for
$J$ and $K$ and of ${\cal P}'_{ap}(J)$ and ${\cal P}'_{ap}(K)$ are equal,
respectively. The values $p'$ and $\Delta'$ are calculated imposing 
that the two lowest moments of ${\cal P}'_{ap}(\Delta)$ match those of the 
real distribution.
This procedure has to be used with some care:
in some systems where the random-field mechanism is important and the
initial randomness is on the interaction ($J$, for instance), forcing the field 
back into a uniform distribution leads to incorrect results.
In Ref. \onlinecite{yeo}, for instance, the crystal-field probability
distribution is maintained uniform throughout the renormalization
procedure. Consequently, the random model critical behavior is 
characteristic of a high-dimensional system: the critical temperature
of the tricritical point diminishes as randomness is increased but only
reaches the zero temperature axis at a finite value of the disorder.
As thouroughly discussed in Ref. \onlinecite{berker}, the mechanism
responsible for the lack of first-order phase transitions in two-dimensional
random systems is the disorder in the crystal-field, which is not taken into 
account by approximations such as the one used in Ref. \onlinecite{yeo}. In
the model we study in this paper, however, the important physical ingredient
is the disorder in the field, which is not approximated away by our RSRG
procedure. Finally, we would like to mention that the way we treated
the random field distribution is not unique.  In this
work we assume that only one field acts on each site and
a weight is associated to the fields (this weight is the ratio between 
the coordination number of the site in the cell and the coordination
number of the site on the square lattice). Conversely, one could also take  the number of fields acting in a given site  as equal to the coordination
number of the site in the cell.  We performed calculations using both procedures above. 
The results do not vary qualitatively (and some times
quantitatively) from one approach to the other. The approach we chose, 
however, leads to simpler expressions, which are easier to deal with 
analitically.

	The expressions connecting renormalized and original parameters
are easily obtained following the procedure outlined above but are
too lenghty to be explicitly written here. Formally, they can be
expressed as
\begin{eqnarray}
J' &=&  J'(p,J,K,\Delta); \: K' = K'(p,J,K,\Delta); \nonumber \\
\Delta' &=& \Delta'(p,J,K,\Delta); \; p' = p'(p,J,K,\Delta) . \label{eq:rgt}
\end{eqnarray}
    Critical points are then evaluated as non-trivial fixed
points of the above relations; phases are identified according to the
attractor of their points. The order of the transition is obtained 
through the study of the largest eigenvalue of the 
renormalization-group transformation (RGT) matrix \cite{fisher}.
More precisely, a first-order phase transition such that 
$m \equiv <S>$ is discontinuous at the transition point is signaled by the 
presence of an eigenvalue equals to $b^d$
in the odd sector of the linearized RGT matrix, where 
$b$ is the length-scaling parameter and $d$ is the dimension of the system. If
$q \equiv <S^2>$ is discontinuous, the $b^d$ eigenvalue belongs 
to the even sector of the RGT matrix.
In the present case, $b^d = N/N' = 5$, where $N$ is the number of bonds 
of the original cell and $N'$ is the number of bonds of the renormalized one.

\section{Results}

	In Fig.~\ref{pure} the pure ($p=0$) phase diagram is depicted, for
completeness. We would like to stress that the dashed line (and its zero
temperature point) is atracted to a fixed point (which depends on the
approximation) where the largest eigenvalue for both
even and odd sections of the RGT matrix equals to $b^d$, indicating a
first order phase transition in $m$ and in $q$. Note that 
the $K=0$ plane is {\it not} an invariant one and the biquadratic 
interaction $K$ is generated by the renormalization transformation.

\begin{figure}
\epsfxsize=6.5cm
\begin{center}
\leavevmode
\epsffile{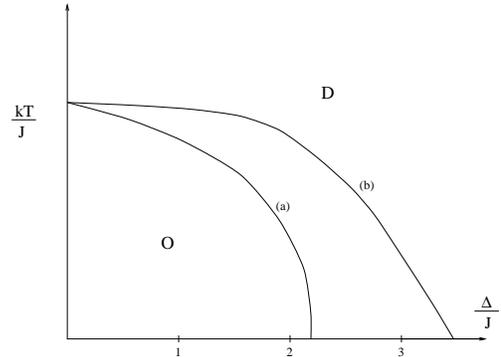}
\caption{Phase diagrams of the RFBC model for $p<p_c$ (see text).
(a) stands for $p=0.1$ and (b) stands for $p=0.3$.  There is no first order
transition and  both continuous lines belong to the pure Ising model
universality class. The frontier at zero temperature is attracted to the
random fixed point (see text).
$O$ ($D$) stands for ordered (disordered) phase.}
\label{random} 
\end{center}
\end{figure}

	Following Ref. \onlinecite{berker}, the first-order transition
should vanish for $p>0$ (random model). This is actually the behavior we 
observe. In fact, the first-order fixed point attractor of the dashed
line in Fig.~\ref{pure} is found to be unstable along the $p$ direction. This
is the expected physical behavior when randomness is introduced. On the
other hand, the attractor of the pure Ising model transition line, namely
$(p^*=0,J^*=0.4407,K^*=-0.0731,\Delta^*=-\infty)$, is stable along the 
same direction. There are still two possibilities for the random 
model critical behavior: either the whole line of continuous transition 
belongs to the 
universality class of the spin-1/2 Ising model or an unstable
fixed point at finite temperature separates the Ising critical line from
another continuous line which belongs to a new universality class. Our
results suport the first option: the Ising critical line extends down to
the zero temperature point (see Fig.~\ref{random}, where typical phase 
diagrams for $p<p_c=1/2$ are depicted). Here there are still two
possible scenarios. The continuous transition for $p \neq 0$ or $1$ belongs
either to the pure or to the disordered Ising model universality class.
For the hierarchical lattice we use in this work, the specific heat
critical exponent of the pure Ising model, $\alpha$, is negative
and disorder is irrelevant, according to the Harris criterion \cite{Har}.
Therefore, the continuous transitions depicted in Figs. 3, 4 and 5 belong
to the pure Ising model universality class. For the corresponding model
on a two-dimensional Bravais lattice, where $\alpha = 0$, the Harris
criterion is inconclusive. The accepted behavior, when disorder is present,
is the following: critical exponents of the random model retain the same
values as their pure conterparts but logarithmic corrections are
introduced by randomness \cite{Fabio}. 
Experimental results also indicate the same critical exponents for
the pure and random two-dimensional Ising model \cite{exper}.
On the other hand, when $\alpha$ is
positive, as in the three-dimensional Ising model, disorder
makes the system to crossover to a new universality class.

Note that the critical value of $\Delta/J$ which separates the ordered and 
disordered phases, $(\Delta/J)_c$,
increases as $p$ decreases, in contrast to the result obtained by
a cluster variational approach on a similar model \cite{mar2}, which
leads to a constant value for $(\Delta/J)_c$ for any $p<p_c$. 
The latter  result might  be an artifact of the cluster variational
approximation.

\begin{figure}
\epsfxsize=6.5cm
\begin{center}
\leavevmode
\epsffile{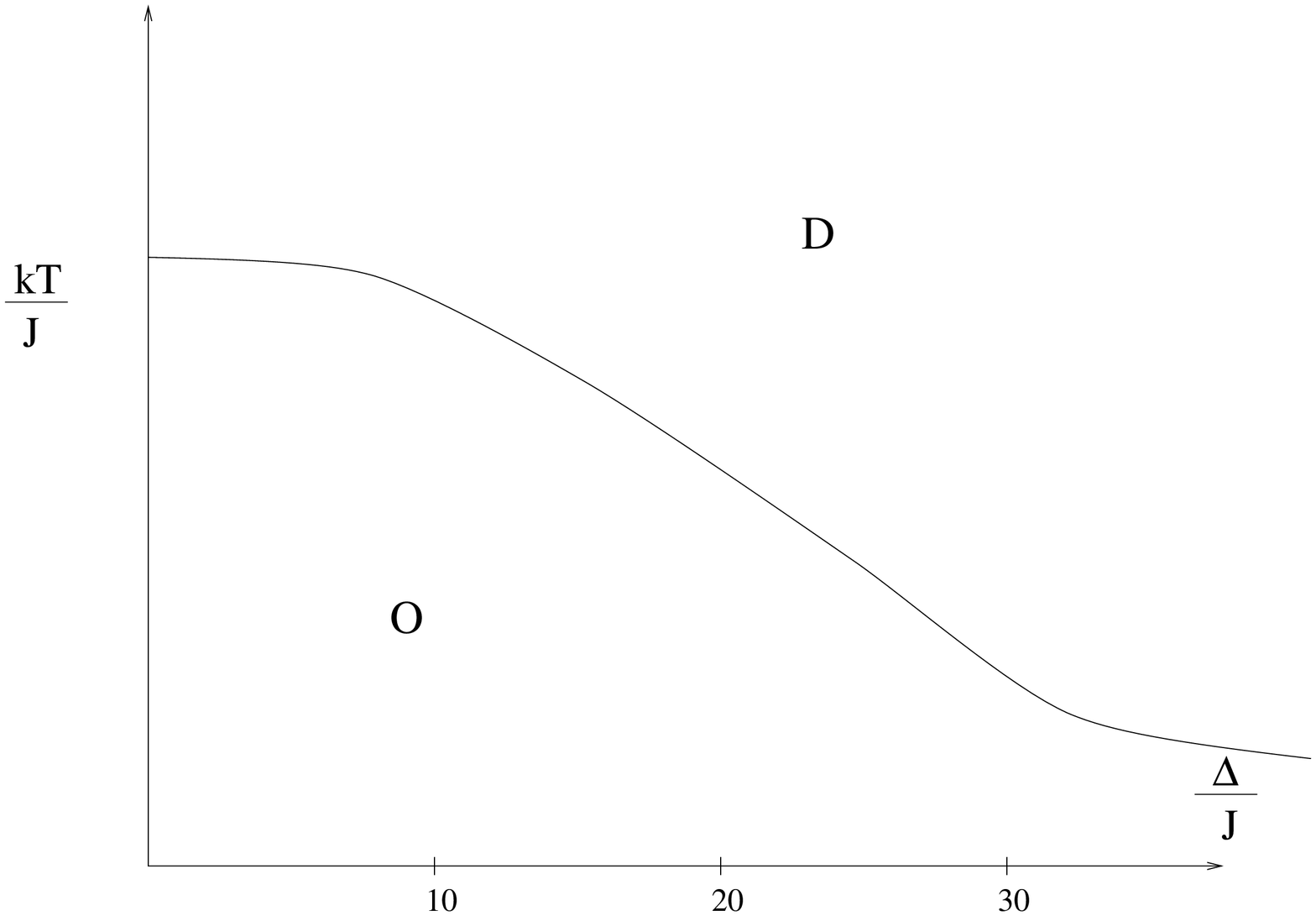}
\caption{Phase diagrams of the RFBC model for $p=1/2$, which is
the value of $p_c$ in our approximation. The critical line touches the
zero temperature axis at $\Delta/J=\infty$.
$O$ ($D$) stands for ordered (disordered) phase.}
\label{pc} 
\end{center}
\end{figure}

	At zero temperature, points on the frontier between the disordered 
and ordered phases flow to a random fixed point, 
$(p^*=1/2,J^*=\infty,K^*=\frac{3}{4} \ln (2) - J^*,\Delta^*=\infty)$,
such that $J^*/\Delta^*=0$. This is the percolation fixed point
of the site-diluted spin-1/2 Ising model. In fact, for $\Delta=\infty$
the RFBC model is equivalent to the random site spin-1/2 Ising model, where
sites are present or absent with probability $p$ or $1-p$ respectively.
This comes from the fact that, for $\Delta=\infty$, a $+\Delta$ crystal
field acting on a given site forces that site to be in the $S=0$ state 
(absent), while a $-\Delta$ field forces the site to be either in the state 
$S=1$ or in the state $S=-1$ (both represent a present site). Thus, only for
high enough $p$ an infinite cluster of $S=\pm 1$ states will form and will be 
able to sustain order. Exactly at $p=p_c$, there is such an infinite
cluster but its critical temperature is zero. Therefore, the critical 
parameter $(\Delta/J)_c$ reaches $\infty$ for $p=p_c$ (see Fig.~\ref{pc}). 
Our evaluation of $p_c$ is 1/2, while the accepted value for the 
site percolation critical probability on the square lattice 
is $p_c=0.5927$ \cite{perco}. It is not
unusual that small-cell RSRG approximations fail to obtain a quantitatively
precise value. Note, however, that we do obtain the correct qualitative
behavior, i.e., a finite value of $p_c$ (contrarily to the standard 
mean-field approximation, which predicts $p_c=0$ \cite{mar1}). 

\begin{figure}
\epsfxsize=6.5cm
\begin{center}
\leavevmode
\epsffile{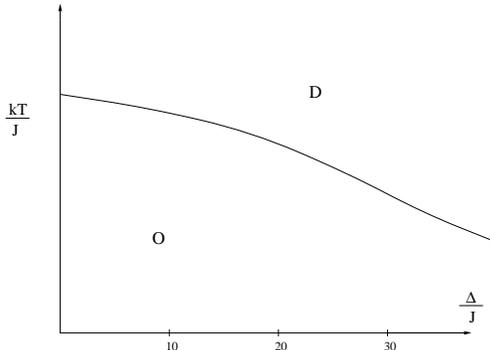}
\caption{Phase diagram of the RFBC model for $p=0.6 > p_c$. 
The critical line never touches the zero temperature axis.
$O$ ($D$) stands for ordered (disordered) phase.}
\label{random1} 
\end{center}
\end{figure}

	For $p>p_c$, the critical line never touches the $(\Delta/J)$
axis. Even at $\Delta/J=\infty$ the infinite cluster of $S=\pm 1$ spins
is, on a large scale, a two-dimensional object and its critical temperature
is finite (see Fig.~\ref{random1}). 

	At this point, it is worthwhile to compare our results with those from mean-field
calculations (see Refs.\ \onlinecite{mar1,mar2,mf1,mf2}). Standard mean-field analysis
leads to a first-order transition inside the ordered phase, ending in a 
critical end point, a reentrant behavior in the $kT/J \times \Delta/J$
diagram and a physically incorrect value for $p_c$. We have already 
commented on this
last feature. Concerning the first-order transition inside the ordered
phase, it has been shown that it is unstable against randomness 
in two dimensions \cite{berker}. Thus, it is expected that a reliable 
approximation to a two-dimensional system will not find such transition. 
Finally, we found no reentrance in our results; actually, in some
other models reentrant behavior has been found for $d=3$ systems, but not in 
their
two dimensional counterpart (see, for example, Ref. \onlinecite{berker2}). 
We should also point out that more sophisticated mean-field-like approximations 
have been 
applied to the RFBC model. They lead to a finite value of
$p_c$ but still predict the existence of first-order transition in the
random model as well as a reentrant behavior. Hence, the results shown 
in this work reflect the correct qualitative behavior of the RFBC model in 
two dimensions. 

\section{Summary}
  
  	A RSRG procedure is applied to the RFBC model in two dimensions.
Our calculation recovers the correct phase diagram of the pure model and
predicts that no first order phase transition is maintained when
randomness is introduced. This is in accordance with general predictions
for two-dimensional disordered models \cite{berker}. We also obtain that
the whole line of continuous transitions, for $p \neq 0$, belongs to the
Ising universality class, discarding the existence of an unstable
fixed point at finite temperature. The zero temperature frontier between 
ordered and
disordered phases, $(\Delta/J)_c$, is attracted to the percolation fixed 
point of the site diluted Ising model. Contrarily to results from a cluster
variational
analysis \cite{mar2}, the value of $(\Delta/J)_c$ increases as $p$ increases.
Such a behavior is also predicted by standard mean-field approximations.

\section{Acknowledgments}

   We would like to thank Prof. P. M. C. de Oliveira, Prof. S. M.
de Oliveira and Prof. S. L. A. de Queiroz for helpful discussions.

\end{document}